\documentclass[12pt]{article}
\usepackage[T1]{fontenc}
\usepackage{geometry}
\usepackage{color}
\usepackage{setspace}
\makeatletter

\providecommand{\LyX}{L\kern-.1667em\lower.25em\hbox{Y}\kern-.125emX\@}
\newcommand{\noun}[1]{\textsc{#1}}
\let\SF@@footnote\footnote
\def\footnote{\ifx\protect\@typeset@protect
    \expandafter\SF@@footnote
  \else
    \expandafter\SF@gobble@opt
  \fi
}
\expandafter\def\csname SF@gobble@opt \endcsname{\@ifnextchar[
  \SF@gobble@twobracket
  \@gobble
}
\edef\SF@gobble@opt{\noexpand\protect
  \expandafter\noexpand\csname SF@gobble@opt \endcsname}
\def\SF@gobble@twobracket[#1]#2{}

 \newcommand{\lyxaddress}[1]{
   \par {\raggedright #1 
   \vspace{1.4em}
   \noindent\par}
 }

\makeatother

\begin{document}

\title{Application of finite field-dependent BRS transformations to problems of the
Coulomb gauge}

\author{Satish D. Joglekar\thanks{
email address: sdj@iitk.ac.in
}}

\maketitle

\lyxaddress{Department of Physics,Indian Institute of Technology,Kanpur;Kanpur 208016 (INDIA)}

\( \qquad \qquad \qquad \qquad \qquad \qquad \qquad \qquad \qquad  \)and

\author{\( \qquad \qquad \qquad \qquad \qquad \qquad  \)Bhabani Prasad Mandal}

\lyxaddress{S.N.BOSE National centre for Basic Sciences, JD Block; Sector-III, Salt-Lake,
Calcutta 7000098 (India)}

\begin{abstract}
We discuss the Coulomb propagator in the formalism developed recently in which
we construct the Coulomb gauge path-integral by correlating it with the well-defined
Lorentz gauge path-integrals through a finite field-dependent BRS transformation.
We discover several features of the Coulomb gauge from it. We find that the
singular Coulomb gauge \emph{has} to be treated as the gauge parameter \( \lambda  \)\( \rightarrow  \)
0 limit. We further find that the propagator so obtained has \emph{good high
energy behavior} (k\( _{0}^{-2} \)) for \( \lambda  \)\( \neq  \)0 and \( \epsilon  \)\( \neq  \)0.
We further find that the behavior of the propagator so obtained is sensitive
to the order of limits k\( _{0} \)\( \rightarrow  \) \( \infty  \), \( \lambda  \)\( \rightarrow  \)
0 and \( \epsilon  \)\( \rightarrow  \) 0 ; so that these have to be handled
carefully in a higher loop calculation. We show that we can arrive at the result
of Cheng and Tsai for the ambiguous two loop Feynman integrals \emph{without}
the need for \emph{}an extra ad hoc regularization and within the path integral
formulation.
\end{abstract}

\section{INTRODUCTION}

The standard model is based upon a nonabelian gauge theory. Various gauge choices
{[}1{]} have been adopted to suit different calculations in Yang-Mills theories.
The most reliable and well established of these are the linear covariant Lorentz
gauges and their spontaneously broken analogs of R\( _{\xi } \)- gauges {[}2{]}.
However, various other choices of gauges have been used for reasons of simplicity
of approach in different contexts. Prominent among these are various non-covariant
gauges : the axial gauges, light-cone gauges, planar gauges and the Coulomb
gauge {[}3, 4{]}. These gauges have been a subject of extensive research {[}3,
4{]} ; however they have not always led to well established and foolproof ways
of calculations as compared with the Lorentz gauges; notwithstanding many attempts.

The major problem in the axial-type gauges is the lack of definition of the
propagator because of \( \frac{1}{(\eta .k)^{n}} \)-type poles in it. A problem
also appears in the Coulomb gauge: the time-like propagator is not damped as
k\( _{0} \)\( ^{-2} \) as k\( _{0} \) \( \rightarrow  \) \( \infty  \)
and hence, unlike the covariant gauges, the k\( _{0} \)-integrals are not as
convergent as those in the Lorentz gauges. Cheng and Tsai {[}5{]} have found
it necessary to introduce an extra regularization to deal with this extra degree
of divergence and have shown additional contributions {[}5,6{]} arising from
this fact which agree with the earlier results based on the Hamiltonian approach
{[}6{]} .Cheng and Tsai have argued {[}5{]} that path-integral formalism is
unable to take care of this problem within its own framework. These questions
of correct definition, in the context of both the axial-type and the Coulomb
gauge, should not for dismissed as unimportant and of peripheral significance;
we do not know how to do an actual calculation until these questions are settled,
not-withstanding any formal approaches and arguments. To see this recall the
following drastic facts (a) a change of prescription for axial poles from Principle
Value prescription to the Leibbrandt-Mandelstam prescription has been known
to alter grossly the behavior of divergence structure even of the one loop integrals
{[}7{]} ;(b) that many different treatments that have been attempted for the
axial pole problem and these have led to different consequences for \emph{observables}
even in low loop orders and thus all of these cannot in particular the compatible
with the well established Lorentz gauges. Thus, unless the problem of the definition
of axial and the Coulomb gauges has been addressed to, any formal treatment
regarding these gauges is, in fact, open to doubt.

Subsequent to the work of Cheng and Tsai, Doust and Doust and Taylor{[}8{]}
have used a type of interpolating gauges to discuss the Feynman integrals in
the Coulomb gauge. Recently, several new attempts have been made at the treatment
of the Coulomb gauge. Baulieu and Zwanziger {[}9{]} have attempted a formal
treatment based on interpolating gauges. We also have the split dimensional
regularization used by Leibbrandt and co-workers {[}10 {]} and NDI{[}11{]} method
has also been applied to the problem. The latter attempts however seem to have
an element of arbitrariness, also present in many axial-gauge treatments, and
further it is not obvious whether the results so obtained will coincide with
those from the Lorentz gauges. 

In this work, we shall attempt a treatment of the Coulomb gauge problem within
the path integral framework itself. We shall show that this framework, is able
to accommodate the Coulomb gauge also. We shall find that no extra regularization,
{[}over and above the usual ultraviolet divergence treatment{]}, is required
to be imposed in our treatment for the ambiguous 2-loop Feynman integrals pointed
out by Cheng and Tsai. We shall be using the path integral framework for connecting
generating functionals of different gauges {[}12,13,14{]} that has been constructed
using finite field dependent BRS {[}FFBRS{]} transformations that have been
employed for the axial gauge problem earlier. In this approach, we start out
with the Lorentz gauge path-integral incorporating the correct \( \epsilon  \)-term,
that takes care of its own poles. We then use an FFBRS transformation that takes
one to the Coulomb gauge {[}14{]}. Such a procedure, apart from preserving the
(correctly defined) vacuum expectation values of gauge-invariant observables,
is expected to take care of all the problems of \emph{definition} of other gauges.
We use this to evaluate the Coulomb gauge propagator and obtain its effective
expression that is simple enough for actual calculations.We hope that this approach
will provide at best a direct route to all problems of the Coulomb gauge and
at least a view complimentary to other approaches {[}5, 6, 8-11{]}.

We shall spell out the plan of the paper. In section 2, we summarize the earlier
results regarding construction of a generating functional for an arbitrary gauge
using the approach of the FFBRS transformations. We also summarize earlier work
regarding the Coulomb gauge. In section 3, we shall evaluate the Coulomb gauge
propagator using the formalism in section 2. In section 4,we study the behavior
of the result under the limits \( \lambda  \)\( \rightarrow  \) 0, \( \epsilon  \)\( \rightarrow  \)
0 and k\( _{0} \)\( \rightarrow  \) \( \infty  \). {[} Here \( \lambda  \)
is the gauge parameter which we need to keep{]} .We show explicitly that there
are terms in the expression of the propagator which are sensitive to the order
of these limits. If we were to take the limits \( \lambda  \)\( \rightarrow  \)
0, \( \epsilon  \)\( \rightarrow  \) 0 first ,we would obtain the {}``ill-behaved{}''
propagator. We show that the high k\( _{0} \)-behavior is normal {[}k\( _{0} \)\( ^{-2} \){]}
if we understand the Coulomb gauge as the \emph{limit} \( \lambda  \)\( \rightarrow  \)
0 to be taken after Feynman integrals are performed. We note that within this
path-integral formulation, there is no obvious necessity to introduce any extra
regularization (unlike the work of Cheng and Tsai.) In section 5, we summarize
our results and offer comparison with other approaches.

\section{PRELIMINARIES}

In this section, we shall introduce our notations and summarize the results
of earlier works of references {[}15,16,12,13{]} on the application of the finite
field-dependent BRS transformations (to axial-type) that we need in this work.We
shall also summarize the problems faced in the calculations using the Coulomb
gauge.

\subsection*{\noun{A. Notations}}

\paragraph{\textmd{The Faddeev-Popov effective action {[}FPEA{]} with a gauge function}
\protect\( \widehat{F^{\gamma }}\protect \)\textmd{{[}A{]} is given by}}

\subparagraph{\textmd{\emph{S\protect\( _{eff}\protect \)}} \textmd{=  S\protect\( _{0}\protect \)
+ S\protect\( _{gf}\protect \) +}\textmd{\emph{S\protect\( _{gh}\protect \)}}
\textmd{~ ~~ ~~ ~~ ~ ~~ ~~ ~~ ~~~ ~~ ~~ ~~ ~~ ~ ~~
~~ ~~ ~~~ ~~ ~~ ~~ ~ ~~ ~~ ~~ ~~ ~~~ ~~~(2.1)}}

with\footnote{%
We understand singular gauges such as Coulomb gauges as the limit \( \lambda  \)
\( \rightarrow  \) 0. We shall see in section 4 the importane of this limiting
procedure.
}

S\( _{gf} \) =  -\( \frac{1}{2\lambda }\int  \) d\( ^{4}x \) \( \widehat{F^{\gamma }} \){[}A{]}\( ^{2} \)
~ ~~ ~~ ~~ ~ ~~ ~~ ~~ ~~~ ~~ ~~ ~~ ~ ~~ ~~ ~~
~~ ~~ ~~ ~~ ~ ~~ ~~ ~~ ~~~ ~~ ~~~~~(2.2a)

and

\emph{S\( _{gh} \) =  -}\( \int  \)d\( ^{4}x \) \( \overline{c} \)\( ^{\alpha } \)\( \widehat{M^{\alpha \beta }}c^{\beta } \)
~ ~~ ~~ ~~ ~ ~~ ~~ ~~ ~~~ ~~ ~~ ~~ ~ ~~ ~~ ~~
~~ ~~ ~~ ~~(2.2b)

with

\( \widehat{M^{\alpha \beta }} \) =  \( \frac{\delta }{\delta }\frac{\widehat{F^{\alpha }[A]}}{A^{\gamma }_{\mu }} \)D\( ^{\gamma \beta } \)\( _{\mu } \)
{[}A{]} ~ ~~ ~~ ~~ ~ ~~ ~~ ~~ ~~~ ~~ ~~ ~~ ~ ~~ ~~
~~ ~~ ~~ ~(2.3)

D\( ^{\alpha \beta } \)\( _{\mu } \) {[}A{]} =  \( \delta ^{\alpha \beta }\partial _{\mu }+g\, f^{\alpha \beta \gamma }A_{\mu }^{\gamma } \)
~ ~~ ~~ ~~ ~ ~~ ~~ ~~ ~~~ ~~ ~~ ~~ ~ ~~ ~~ ~~
~~ ~~(2.3a)

\paragraph{\textmd{BRS transformations for the three effective actions are:}}

\subparagraph{\textmd{\protect\( \phi \protect \)'\protect\( _{i}\protect \) =  \protect\( \phi \protect \)\protect\( _{i}\protect \)
+\protect\( \delta _{iBRS}[\phi ]\delta \Lambda \protect \)~ ~~ ~~ ~~
~ ~~ ~~ ~~ ~~~ ~~ ~~ ~~ ~~ ~~~ ~~ ~~ ~~ ~ ~~ ~~
~~ ~~ (2.4)}}

\subparagraph{\textmd{with \protect\( \delta _{iBRS}[\phi ]\protect \) equal to D\protect\( ^{\alpha \beta }\protect \)\protect\( _{\mu }\protect \)c\protect\( ^{\beta }\protect \),-1/2
g\protect\( f^{\alpha \beta \gamma }\protect \)c\protect\( ^{\beta }\protect \)c\protect\( ^{\gamma }\protect \),
and \protect\( \widehat{F}/\lambda \protect \) respectively for A, c and \protect\( \overline{c}\protect \).}}

\subparagraph{\textmd{We consider two arbitrary gauge fixing functions F{[}A{]} and F'{[}A{]}
which could be nonlinear or non-covariant and an interpolating gauge function }}

\subparagraph{\textmd{F\protect\( ^{M}[A]\protect \) =  \protect\( \kappa \protect \)F'{[}A{]}+(1-\protect\( \kappa \protect \))F{[}A{]};
.~~~0\protect\( \leq \kappa \leq 1\protect \).~~ ~~ ~~ ~~~ ~~
~~ ~~~ ~~ ~~ ~~ ~(2.5)}}

\subparagraph{\textmd{We denote the FPEA for the three cases \protect\( \widehat{F}\protect \)
= F,F' and F\protect\( ^{M}\protect \) by} \textmd{\emph{S\protect\( _{eff}\protect \),
S \protect\( '_{eff}\protect \) and S\protect\( ^{M}\protect \)\protect\( _{eff}\protect \)}}
\textmd{respectively. In the case of the mixed gauge condition, \protect\( \delta _{iBRS}[\phi ]\protect \)
for \protect\( \overline{c}\protect \) is \protect\( \kappa \protect \)-dependent
and in this case we show this explicitly by expressing (2.4) as}}

\subparagraph{\textmd{\protect\( \phi \protect \)'\protect\( _{i}\protect \) =  \protect\( \phi \protect \)\protect\( _{i}\protect \)
+\protect\( \widetilde{\delta }_{iBRS}[\phi ,\kappa ]\delta \Lambda \protect \)\protect\( \equiv \phi \protect \)\protect\( _{i}\protect \)
+\protect\( \{\widetilde{\delta _{1}}_{iBRS}[\phi ]+\kappa \protect \)\protect\( \widetilde{\delta _{2}}_{iBRS}[\phi ]\}\delta \Lambda \protect \)~
~~ ~~ ~~ ~~ (2.6)}}

\subparagraph{\textmd{In this work, we shall be interested in} \textmd{\emph{arriving at
the correct definition of the path integral for the Coulomb gauge}} \textmd{by
correlating it to the well-defined path integral Lorentz gauges by the procedure
similar to that in References {[}12,13{]} and deducing the results from it particularly
how the Coulomb propagator should be handled. We therefore choose F =  \protect\( \partial .A\protect \)
, F' =  -\protect\( \nabla \protect \).}A \protect\( \equiv \protect \) -
\textmd{\protect\( \sum \protect \) \protect\( \partial _{i}A_{i}\protect \)
and}\footnote{%
The minus sign in F' is a matter convenience and not of necessity.
} \textmd{}}

\subparagraph{\textmd{F\protect\( ^{M}\protect \) = \protect\( -\kappa \protect \)\protect\( \nabla \protect \).}A\textmd{+(1-\protect\( \kappa \protect \))\protect\( \partial .A\protect \)
=\protect\( \partial .A\protect \)-\protect\( \kappa \protect \)\protect\( \partial _{0}A\protect \)\protect\( _{0}\protect \)~~
~~ ~~ ~~~ ~~ ~~~~~ ~~ ~~ ~~~(2.7)}}

\subparagraph{\noun{B.Earlier Results using FFBRS}}

\paragraph{\textmd{Following observations in {[}14{]} and {[}16{]}, we know the the field
transformations that takes one from the path integral in a gauge F to that in
another gauge F'. It is given {[}15{]} by the finite field-dependent BRS {[}FFBRS{]}
transformation }}

\subparagraph{\textmd{\protect\( \phi \protect \)'\protect\( _{i}\protect \) = \protect\( \phi \protect \)\protect\( _{i}\protect \)
+\protect\( \delta _{iBRS}[\phi ]\Theta \protect \) {[}\protect\( \phi ]\protect \)
~ ~~ ~~ ~~ ~ ~~ ~~ ~~ ~~~ ~~ ~~ ~~ ~ ~~ ~~ ~~
~~ ~(2.8)}}

\subparagraph{\textmd{where \protect\( \Theta [\phi \protect \){]} has been constructed
by the integration {[}15{]} of the infinitesimal field-dependent BRS {[}IFBRS{]}
transformation}}

\subparagraph{\protect\( \frac{d\phi _{i}}{d\kappa }=\protect \) \protect\( \delta _{iBRS}[\phi (\kappa )]\Theta \protect \)\textmd{'}{[}\protect\( \phi \protect \)(\protect\( \kappa \protect \)){]}~
~~ ~~ ~~ ~ ~~ ~~ ~~ ~~~ ~ ~~ ~~ ~~ ~~ ~ ~~ ~~
~ ~\textmd{(2.9)}}

(where \( \delta _{iBRS}[\phi (\kappa )] \) refers to the BRS variations of
the gauge F with the boundary condition \( \phi  \){[}\( \kappa  \)=1{]}=\( \phi  \)'and
\( \phi  \){[}\( \kappa  \)=0{]}=\( \phi  \) and is given in a closed form
by {[}15{]}

\( \Theta  \){[}\( \phi  \){]} = \( \Theta  \)'{[}\( \phi  \){]}{[}exp\{f{[}\( \phi  \){]}\}-1{]}/f{[}\( \phi  \){]}~
~~ ~~ ~~ ~ ~~ ~~ ~~ ~~~ ~~ ~~ ~~ ~ ~~ ~(2.10)

and f is given by

f = \( \sum _{i} \) \( \frac{\delta \Theta '}{\delta \phi _{i}} \)\( \delta _{iBRS}[\phi ] \)~~
~~ ~~ ~ ~~ ~~ ~~ ~~~ ~~ ~~ ~~ ~ ~~ ~~ ~~ ~~ ~~
~(2.10a)

In the present case, \( \Theta  \)' is given by {[}14 {]} ,

\( \Theta  \)'{[}\( \phi  \)(\( \kappa  \)){]} = \( i\int d^{4}y \) \( \overline{c} \)\( ^{\gamma } \)(y)
(F\( ^{\gamma } \){[}A(\( \kappa )] \) -F' \( ^{\gamma } \){[}A(\( \kappa )] \)) 

~ ~~ ~~ ~~ = \( i\int d^{4}y \) \( \overline{c} \)\( ^{\gamma } \)(y)\( \partial _{0}A \)\( _{0} \)
~ ~~ ~~ ~~ ~~ ~ ~~ ~~ ~~ ~~ ~(2.11)

We also note,

\emph{S\( _{eff} \)\( ^{M}[ \)}\( \phi  \)(\( \kappa  \)),\( \kappa ] \)\( \equiv  \)S\( _{0} \)~-\( \frac{1}{2\lambda }\int  \)
d\( ^{4}x \) {[}\( \partial .A \)-\( \kappa  \)\( \partial _{0}A \)\( _{0} \){]}\( ^{2} \)
\emph{-}\( \int  \)d\( ^{4}x \) \( \overline{c} \)\( ^{\alpha } \)\( [\partial ^{\mu }D_{\mu }-\kappa \partial ^{0}D_{0}]^{a\beta }c^{\beta } \)~~
~~ ~ ~~ ~~ ~~ ~~ ~~ ~(2.12) 

Using this type of transformation, a result had been established connecting
Green's functions in a pair of gauges F and F' {[}12,14,13, see also 17{]}.
We shall first recapitulate the philosophy behind the method used that was laid
out in Ref. 12,13 and 14 {[}See ref.19 for a summary of the present approach{]}.
We \emph{require} this formulation to be used, because the naive path-integrals
in noncovariant gauges are, a priori, ill-defined and and this formalism affords
a way for constructing a well-defined path-integral for these gauges by correlating
them to the corresponding well-defined path-integral for the Lorentz gauges.
The proper definition of the Green's functions in Lorentz gauges is not possible
till we include a term -i\( \epsilon  \)\( \int  \)d\( ^{4}x \)(\( \frac{1}{2} \)AA
-\( \overline{c} \) c) in the effective action that says \emph{how} its \emph{poles should be treated.}
Similarly, the proper definition of the Coulomb Green's functions in gauges
requires that we include an appropriate \( \epsilon  \)-term and this is expected
to tell us how the Coulomb propagator should be handled.The correct \( \epsilon  \)-term
for the Coulomb gauges, in particular, is automatically obtained by the FFBRS
transformation of that connects the path integrals in the Lorentz and the Coulomb
gauges as was done in {[}12,13{]} for the axial case. It was shown {[}14{]}
that the effect of this term on the Coulomb gauge Green's functions is expressed
in the simplest form when it is expressed in the relation (2.13 ) below and
happens to be simply to add to \emph{S\( _{eff} \)\( ^{M}[ \)}\( \phi  \),\( \kappa ] \)
the same -i\( \epsilon  \)\( \int  \)d\( ^{4}x \)(\( \frac{1}{2} \)AA -\( \overline{c} \)
c) \emph{inside the \( \kappa  \)-integral.} Thus, taking care of the proper
definition of the Coulomb Green's functions, these Green's functions, \emph{compatible
with those in Lorentz gauges,} are given by the result in {[}13,14{]} which
in the present context reads,

<\textcompwordmark{}<O{[}\( \phi  \){]}>\textcompwordmark{}>\( _{coul} \)
=  <\textcompwordmark{}<O{[}\( \phi  \){]}>\textcompwordmark{}>\( _{L} \) 

~~~ ~~ ~ ~~ ~~ ~~ ~~+i\( \int  \)\( _{_{0}} \)\( ^{1} \)d\( \kappa  \)
\( \int  \)D\( \phi  \) exp\{ i \emph{S\( _{eff} \)\( ^{M}[ \)}\( \phi  \),\( \kappa ] \)+\( \int  \)d\( ^{4} \)z
\( \varepsilon [\frac{1}{2} \)A\( _{\mu } \)A\( ^{\mu } \)- \( \overline{c} \)c
{]}\}

~~~~ ~~ ~ ~~ ~~~~ ~~ ~\( \bullet  \)\( \sum _{i}(\widetilde{\delta _{1i}} \)
{[}\( \phi  \){]}+\( \kappa  \)\( \widetilde{\delta _{2i}} \){[}\( \phi  \){]})(-i\( \Theta ') \)\( \frac{\delta ^{L}O}{\delta \phi _{i}} \)~~
~ ~~~ ~~ ~ ~~ ~~~ ~~(2.13)

When applied to the two-point function,\{O{[}\( \phi  \){]} =  A\( ^{\alpha }_{\mu }(x) \)A\( ^{\beta } \)\( _{\nu }(y)\} \),
in obvious notations, (2.13 ) reads {[}13{]},

iG\( ^{C\alpha \beta } \)\( _{\mu \nu } \)(x-y) = iG\( ^{L\alpha \beta } \)\( _{\mu \nu } \)(x-y)

~ ~~~ ~~ ~ ~~+i\( \int  \)\( _{_{0}} \)\( ^{1} \)d\( \kappa  \)
\( \int  \)D\( \phi  \) exp\{ i \emph{S\( _{eff} \)\( ^{M}[ \)}\( \phi  \),\( \kappa ] \)
+\( \epsilon  \)\( \int  \)d\( ^{4}z \)(\( \frac{1}{2} \)AA -\( \overline{c} \)
c)\}

~~ ~~ ~ ~~ ~~ ~~ ~\( \bullet [ \)D\( _{\mu } \)c\( ^{\alpha } \)(x)A\( ^{\beta } \)\( _{\nu }(y) \)+A\( ^{\alpha }_{\mu }(x) \)D\( _{\mu } \)c\( ^{\beta } \)(y){]}
\( \int d^{4}z \) \( \overline{c} \)\( ^{\gamma } \)(z) {[}\( \partial _{0} \)A\( ^{0\gamma } \)(\( z) \)
{]}~~ ~~~ ~~~~ (2.14)

The above relation gives the value of the exact Coulomb propagator \emph{compatible
to the Green's functions in Lorentz gauges .}The result is exact to all orders.
As mentioned in Ref. {[}12{]}, to any finite order in g, the right hand side
can be evaluated by a finite sum of Feynman diagrams. 

We can consider the above relation to O{[}g\( ^{0}] \).Then it will give the
Coulomb propagator compatible with the Lorentz gauges.\emph{This is automatically
expected to give information as to how the problems associated in the Coulomb
gauge propagator should be dealt with.}It reads,

iG\( ^{0C\alpha \beta } \)\( _{\mu \nu } \)(x-y) = iG\( ^{0L\alpha \beta } \)\( _{\mu \nu } \)(x-y)+i\( \int  \)\( _{_{0}} \)\( ^{1} \)d\( \kappa  \)
\( \int  \)D\( \phi  \) exp\{ i \emph{S\( _{eff} \)\( ^{M}[ \)}\( \phi  \),\( \kappa ] \)
+\( \epsilon  \)\( \int  \)d\( ^{4}z \)(\( \frac{1}{2} \)AA -\( \overline{c} \)
c)\}

{}~~ ~ ~~~ ~~ ~ ~~ ~~ ~ ~~ ~~ ~~ ~\( \bullet [ \)\( \partial  \)\( _{\mu } \)c\( ^{\alpha } \)(x)A\( ^{\beta } \)\( _{\nu }(y) \)+A\( ^{\alpha }_{\mu }(x) \)\( \partial _{\nu } \)c\( ^{\beta } \)(y){]}
\( \int d^{4}z \) \( \overline{c} \)\( ^{\gamma } \)(z) {[}\( \partial _{0} \)A\( ^{0} \)\( ^{\gamma } \)(\( z) \){]}~~
~~~~ (2.15)

{}We, of course, need to evaluate the last term to O{[}g\( ^{0}] \). G\( ^{0} \)
refers to the free propagator. Evaluating this to this order,the above relation,
in momentum space, reads 

iG\( ^{0C} \)\( _{\mu \nu } \)(\( k \)) = iG\( ^{0L} \)\( _{\mu \nu } \)(\( k \))+i\( \int  \)\( _{_{0}} \)\( ^{1} \)d\( \kappa  \){[}\( k_{\mu } \)G\( ^{0M} \)(k)\( k^{0} \)G\( _{0\nu }^{0M} \)+
(\( \mu  \),k)\( \leftrightarrow  \)(\( \nu  \),-k){]}~~ ~~~~ (2.16)

Here, G\( ^{0M} \)(k) and G\( _{0\nu }^{0M} \) are the mixed gauge propagators
for the ghost and the gauge fields derived from \emph{\{S\( _{eff} \)\( ^{M}[ \)}\( \phi  \),\( \kappa ] \)
-i\( \epsilon  \)\( \int  \)d\( ^{4}z \) (\( \frac{1}{2} \)AA -\( \overline{c} \)
c)\}{[}13{]}.

We shall evaluate this in section 3.

\textbf{C:} \textbf{\noun{Problems with the Coulomb gauge}}

The problems associated with the Coulomb gauge in the Lagrangian framework have
been discussed by Cheng and Tsai {[}5,6{]} and in more details by Doust and
Doust and Taylor {[}8{]}. It arises from the fact that the timelike propagator
D\( _{00} \)(k) 

D\( _{00} \)(k) = \( \frac{1}{|\mathbf{k}|^{2}} \)

is not damped as k\( _{0} \) \( \rightarrow  \) \( \infty  \) unlike the
Feynman gauge propagator ,which for all components goes like k\( _{0} \)\( ^{-2} \).
As a result, the k\( _{0} \)-integrals are no longer as convergent as the corresponding
ones in the Feynman gauge. Thus, they require a special treatment.Cheng and
Tsai {[}5,6{]} have provided a treatment in the form of an extra regularization
put in by hand and arrived at nontrivial results for the 2 loop diagrams that
are consistent with the extra terms from the Hamiltonian treatment {[}6{]}.
Doust {[}8{]} provided a treatment based on an interpolating gauge with an interpolating
parameter \( \theta  \).The procedure provides an intrinsic regularization
in terms of the parameter \( \theta  \) and the Coulomb gauge results are understood
as the limit \( \theta  \) \( \rightarrow  \) 0. Doust and Taylor {[}8{]}
have pointed out ,however, difficulties that arise in three loop order.

\section{\noun{Evaluation of the Coulomb propagator }}

In this section,we shall briefly present the evaluation of the propagator of
Eq.(2.16). We shall proceed mainly along the lines of references {[}13,18{]}
and rely on several algebraic results there.

We recall that in (2.16), G\( ^{0M} \)(k,\( \kappa  \)) the ghost propagator
in the mixed gauge, obtained from the quadratic form in \{ S\( ^{M}_{eff} \)
-i \( \epsilon  \)\( \int  \)d\( ^{4}x \)(\( \frac{1}{2} \)AA -\( \overline{c} \)
c)\}. It is given by 

\emph{G\( ^{0M} \) = \( - \)\( \frac{1}{-q^{2}+\kappa q_{0}^{2}-i\varepsilon } \)}
~ ~ ~~~ ~~ ~ ~~ ~~ ~~ ~~~ ~~(3.1)

On the other hand, G\( _{\mu \nu } \)\( ^{0M} \)(k,\( \kappa  \)) is the
mixed gauge propagator for the gauge boson obtained similarly by inverting the
quadratic form Z\( _{\mu \nu } \) in \{ S\( ^{M}_{eff} \) -i \( \epsilon  \)\( \int  \)d\( ^{4}x \)(\( \frac{1}{2} \)AA
-\( \overline{c} \) c)\}. Z\( _{\mu \nu } \) is given, with obvious conventions,
by 

Z\( _{ij} \) =  -g\( _{ij} \)(k\( ^{2} \)+i\( \epsilon  \)) + \( \frac{\lambda -1}{\lambda } \)k\( _{i} \)k\( _{j} \)
;~ ~ ~~ ~~ ~~ ~~~ ~~(3.2a)

Z\( _{0i} \) = Z\( _{i0} \) =  k\( _{i} \)k\( _{0} \) {[}1-\( \frac{(1-\kappa )}{\lambda } \){]}
; ~ ~ ~~ ~~ ~~ ~~~ ~~(3.2b)

Z\( _{00} \) = -g\( _{00} \)(k\( ^{2} \)+i\( \epsilon  \)) + k\( _{0} \)\( ^{2} \)\{1-\( \frac{(1-\kappa )^{2}}{\lambda } \)\}~
~~ ~~ ~~ ~~~ ~~(3.2c)

and G\( _{\mu \nu } \)\( ^{0M} \)(k,\( \kappa  \)) =  Z\( ^{-1}_{\mu \nu } \)
is given by,

Z\( ^{-1}_{00} \) =  \( \frac{-1}{k^{2}+i\varepsilon } \)\( \left\{ \frac{k^{2}+i\varepsilon \lambda +(\lambda -1)k_{0}^{2}}{k^{2}+i\varepsilon \lambda +k_{0}^{2}\kappa (\kappa -2)+\frac{k_{0}^{2}|\mathbf{k}|^{2}\kappa ^{2}}{k^{2}+i\varepsilon }}\right\}  \);~
~~ ~~ ~~ ~~(3.3a)

Z\( ^{-1}_{0i} \) = Z\( ^{-1}_{i0} \) =  \( \frac{-1}{k^{2}+i\varepsilon } \)\( \left\{ \frac{(\lambda -1+\kappa )k_{0}k_{i}}{k^{2}+i\varepsilon \lambda +k_{0}^{2}\kappa (\kappa -2)+\frac{k_{0}^{2}|\mathbf{k}|^{2}\kappa ^{2}}{k^{2}+i\varepsilon }}\right\}  \);~
~~ ~~ ~~ ~~~ ~~(3.3b)

Z\( ^{-1}_{ij} \) =  \( \frac{-1}{k^{2}+i\varepsilon } \)\( \left\{ g_{ij}+k_{i}k_{j}\frac{[(\lambda -1)\{k^{2}+i\varepsilon \lambda +k_{0}^{2}\kappa (\kappa -2)+\frac{k_{0}^{2}|\mathbf{k}|^{2}\kappa ^{2}}{k^{2}+i\varepsilon }\}(\lambda -1+\kappa )^{2}k_{0}^{2}]}{[k^{2}+i\varepsilon \lambda +(\lambda -1)k_{0}^{2}][k^{2}+i\varepsilon \lambda +k_{0}^{2}\kappa (\kappa -2)+\frac{k_{0}^{2}|\mathbf{k}|^{2}\kappa ^{2}}{k^{2}+i\varepsilon }]}\right\}  \)~~
~~~ ~~(3.3c)

We use these in eq.(2.16) to evaluate the gauge boson propagator.We shall keep
\( \lambda  \)\( \neq  \)0 as it will turn out to be necessary. {[}See section
4{]}

The problems associated with the Coulomb gauge arise from the fact that the
naive propagator for the time-like component behaves as k\( _{0} \)\( ^{0} \)
rather than k\( _{0} \)\( ^{-2} \). The transverse components do not cause
this problem. Hence, we focus attention on G\( _{00} \)\( ^{0C} \)(k). We
shall find that we can deduce several conclusions on how the Coulomb gauge should
be handled from this calculation alone.We shall therefore evaluate G\( _{\mu \nu } \)\( ^{0C} \)(k)
for \( \mu  \) = \( \nu  \) = 0. We shall do this by following a procedure
as in reference {[}13{]}.We note that in the \( \kappa  \)-integral term:

(i) There is a common denominator linear in \( \kappa  \) coming from G\( _{\mu \nu } \)\( ^{0M} \)(k,\( \kappa  \)).
We express it as (\( \kappa  \)-a\( _{1} \)) k\( ^{2}_{0} \) where a\( _{1} \)
\( \equiv  \)\( \frac{k^{2}+i\varepsilon }{k^{2}_{0}} \).

(ii) There is a factor, quadratic in \( \kappa  \), in the denominator in the
integrand coming from G\( _{00} \)\( ^{0M} \)(k,\( \kappa  \)).We express
this as 

(\( \kappa  \)-\( \kappa  \)\( _{1} \))(\( \kappa  \)-\( \kappa  \)\( _{2} \))k\( ^{2}_{0} \)
\{1+\( \frac{|\mathbf{k}|^{2}}{k^{2}+i\varepsilon } \)\} = (\( \kappa  \)-\( \kappa  \)\( _{1} \))(\( \kappa  \)-\( \kappa  \)\( _{2} \))\( \frac{k_{0}^{2}(k_{0}^{2}+i\varepsilon )}{k^{2}+i\varepsilon } \) 

(iii) The numerator in the integrand is a constant; but in order to use the
existing results in reference 13, we shall express it as k\( ^{2}_{0} \) \textbf{|k|\( ^{2} \)\{(}\( \kappa  \)+\( \alpha  \))-\( \kappa  \)\}.

We thus have 

G\( ^{0C} \)\( _{00} \)(\( k \))-G\( ^{0L} \)\( _{00} \)(\( k \)) =  \textcolor{magenta}{-}\( \frac{4|\mathbf{k}|^{2}}{k^{2}_{0}(k^{2}_{0}+i\varepsilon )} \)\{
I (\( \alpha  \))- I(0)\} ~~ ~~ ~~~ ~~(3.4)

with

I (\( \alpha  \)) \( \equiv  \)\( \int  \)\( ^{^{1}}_{_{0}} \) d\( \kappa  \)
\( \frac{\kappa +\alpha }{(\kappa -a_{1})(\kappa -\kappa _{1})(\kappa -\kappa _{2})} \)
~~ ~~ ~~~ ~~(3.4a)

Here, we note the values 

\( \alpha  \) =  - \( \frac{1}{2} \)\( \frac{\lambda k_{0}^{2}-|\mathbf{k}|^{2}+i\varepsilon \lambda }{|\mathbf{k}|^{2}} \)~
~~ ~~ ~~~ ~~~~ ~~(3.5a)

a\( _{1} \) = \( \frac{k^{2}+i\varepsilon }{k^{2}_{0}} \); \( \kappa  \)\( _{1,2} \)
= \( \frac{k^{2}+i\varepsilon }{k^{2}_{0}+i\varepsilon } \)\( \pm  \)\( \Delta  \)
\( \equiv  \) \( \kappa  \)\( _{0} \)\( \pm  \)\( \Delta  \);~ ~ ~~
~~ ~~ ~~~ ~~(3.5b)

\( \Delta  \) \( \equiv \sqrt{Y} \); \( Y=i\varepsilon  \)\( \frac{k^{2}+i\varepsilon }{(k^{2}_{0}+i\varepsilon )^{2}k^{2}_{0}} \)\{|\textbf{k|\( ^{2} \)-\( \lambda  \)(\( k^{2}_{0} \)+}i\textbf{\( \varepsilon  \))\}}
~~ ~~ ~~ ~~~ ~~(3.5c)

Here, \( \kappa  \)\( _{1,2} \) are the roots of the quadratic form appearing
in the denominator of G\( _{00} \)\( ^{0M} \)(k,\( \kappa  \)).As in ref.
{[}12{]}, we shall find it is convenient to choose a definition of 1/\( \Delta  \)
and we shall choose the following definition for the quantity 1/\( \Delta  \)\( \equiv 1/\sqrt{Y} \)
{[}though we are at liberty to choose other possibilities which will only alter
which root is which{]}. We shall assume that

1/\( \Delta  \)\( \equiv  \)\( \frac{k_{0}(k_{0}^{2}+i\varepsilon )}{\sqrt{i\varepsilon }} \)\( \frac{1}{\sqrt{(k^{2}+i\varepsilon )\{|\mathbf{k}|^{2}-\lambda (k_{0}^{2}+i\varepsilon )\}}} \)

with the branch-cuts ( in the complex k\( _{0} \)-plane) for \( \sqrt{k^{2}+i\varepsilon } \)
and \( \sqrt{|\mathbf{k}|^{2}-\lambda (k_{0}^{2}+i\varepsilon }) \) being taken
along the infinite lines passing through \( \sqrt{|\mathbf{k}|^{2}-i\varepsilon } \)
and -\( \sqrt{|\mathbf{k}|^{2}-i\varepsilon } \) in the first case and \( \sqrt{|\mathbf{k}|^{2}/\sqrt{\lambda }-i\varepsilon } \)
and -\( \sqrt{|\mathbf{k}|^{2}/\sqrt{\lambda }-i\varepsilon } \) in the second
case, with the (finite) segments joining two branch-points deleted. Moreover,
for a real k\( _{0} \) , such that \( k^{2} \)>\textcompwordmark{}> \( \epsilon  \)
and |\textbf{k|\( ^{2} \)-}\( \lambda  \)\( k_{0}^{2} \) >\textcompwordmark{}>
0; we define the phase of \( \sqrt{k^{2}+i\varepsilon } \) \( \sqrt{|\mathbf{k}|^{2}-\lambda (k_{0}^{2}+i\varepsilon }) \)
to be \textasciitilde{} 0. With these definitions, \( \Delta  \) is an odd
function of k\( _{0} \) for -\( \infty  \) < k\( _{0} \) < \( \infty  \).
We shall use the properties of \( \Delta  \) arising from this choice in Appendix
B.

The integral in (3.4 ) is precisely the same as the one occurring in Ref.{[}13{]}
{[} See eq.(60) there{]} in the treatment of the axial gauges. We recall the
result in eq.(61) of Ref.{[}13{]}:

\( \int  \)\( ^{^{1}}_{_{0}} \) d\( \kappa  \) \( \frac{\kappa +\alpha }{(\kappa -a_{1})(\kappa -\kappa _{1})(\kappa -\kappa _{2})} \) 

= \( \frac{1}{(a_{1}-\kappa _{2})} \)\{\( \frac{a_{1}+\alpha }{(-\kappa _{1}+a_{1})} \)\( ln \){[}\( \frac{\kappa _{1}(1-a_{1})}{a_{1}(1-\kappa _{1})} \){]}-\( \frac{\kappa _{2}+\alpha }{\kappa _{1}-\kappa _{2}} \)\( ln \){[}\( \frac{\kappa _{2}(1-\kappa _{1})}{\kappa _{1}(1-\kappa _{2})} \){]}\}
~~ ~~ ~~ ~~~ ~~(3.6)

We shall find convenient to restructure the above equation as,

= \( \frac{1}{D} \)\{\( \frac{(a_{1}+\alpha )}{2} \)\( ln \){[}\( \frac{\kappa _{1}\kappa _{2}(1-a_{1})^{2}}{a^{2}_{1}(1-\kappa _{1})(1-\kappa _{2})} \){]}+\( \frac{1}{2\Delta } \){[}(a\( _{1} \)-\( \kappa  \)\( _{0} \))(\( \alpha  \)+\( \kappa  \)\( _{0} \))
+ \( \Delta  \)\( ^{2} \){]}\( ln \){[}\( \frac{\kappa _{1}(1-\kappa _{2})}{\kappa _{2}(1-\kappa _{1})} \){]}\}
~ ~~ ~~ ~~~ ~~(3.7)

where 

D\( \equiv  \) (a\( _{1}-\kappa _{1}) \)(a\( _{1}-\kappa _{2}) \)~~ ~~
~~~~ ~~~~~~~~ ~~~~ ~~~~~~~ ~~~ ~~(3.7a)

This leads us to 

I (\( \alpha  \))- I(0) = \( \frac{1}{D} \)\{\( \frac{\alpha }{2} \)\( ln \){[}\( \frac{\kappa _{1}\kappa _{2}(1-a_{1})^{2}}{a^{2}_{1}(1-\kappa _{1})(1-\kappa _{2})} \){]}+\( \frac{1}{2\Delta } \){[}(a\( _{1} \)-\( \kappa  \)\( _{0} \))\( \alpha  \)
{]}\( ln \){[}\( \frac{\kappa _{1}(1-\kappa _{2})}{\kappa _{2}(1-\kappa _{1})} \){]}\}
~ ~~ ~~ ~~~ ~~(3.7b)

We then obtain,

G\( ^{0C} \)\( _{00} \)(\( k \))-G\( ^{0L} \)\( _{00} \)(\( k \)) = -
\( \frac{2|\mathbf{k}|^{2}\alpha }{k^{2}_{0}(k^{2}_{0}+i\varepsilon )} \)\( \frac{1}{D} \)\{{[}-\( ln \){[}\( \frac{(1-\kappa _{0})^{2}-\Delta ^{2}}{(1-a_{1})^{2}} \){]}

~~ ~~~ ~~~ ~~~ ~~ + \( ln \){[}\( \frac{\kappa ^{2}_{0}-\Delta ^{2}}{a^{2}_{1}} \){]}{]}+\( \frac{1}{\Delta } \)(a\( _{1} \)-\( \kappa  \)\( _{0} \))\( ln \){[}1+
\( \frac{2\Delta }{\kappa _{2}(1-\kappa _{1})} \){]}\} ~~~~ ~~~~~
~~ ~~~ ~~(3.8)

The above propagator as a function of \( \lambda  \),\( \epsilon  \),and k\( _{0} \)
is admittedly a complicated function and we shall give an effective simpler
treatment for it in Sec 4.But as we shall see in the next section, it reveals
much about the way the Coulomb gauge should be handled.

For completeness, we state the result for the other components of 

\{G\( ^{0C\alpha \beta } \)\( _{\mu \nu } \)(\( k \))-G\( ^{0L\alpha \beta } \)\( _{\mu \nu } \)(\( k \))\}.These
involve,

\( k^{0} \)\( k_{\mu } \)\( \int  \)\( _{_{0}} \)\( ^{1} \)d\( \kappa  \)G\( ^{0M} \)(k,\( \kappa  \))G\( _{0\nu }^{0M} \)(k,\( \kappa  \)) 

For \( \nu  \) = 0, we have already evaluated this.For \( \nu  \) = i, this
becomes,

\( \frac{k_{0}k_{\mu }}{k_{0}^{4}(k_{0}^{2}+i\varepsilon )} \)\( \int  \)\( _{_{0}} \)\( ^{1} \)d\( \kappa  \)
\( \frac{(\kappa +\lambda -1)k_{0}k_{i}}{(\kappa -a_{1})(\kappa -\kappa _{1})(\kappa -\kappa _{2})} \)
= \( \frac{k_{i}k_{\mu }}{k_{0}^{2}(k_{0}^{2}+i\varepsilon )} \)I(\( \lambda  \)-1)
~~ ~~ ~~~~ ~~~~~ ~~ ~~~ ~~(3.9)

I(\( \alpha  \)) has been dealt with in the next section in detail; it needs
to be used in (3.9).

\section{PROPERTIES OF THE PROPAGATOR}

In this section , we shall establish several unusual facets of the Coulomb propagator
we have derived. These essentially depend on the behavior of the propagator
as a function of \( \lambda  \), \( \epsilon  \) and k\( _{0} \) \emph{and
the order in which various limits involving these variables are taken.} In particular,
we shall study the order of \( \lambda  \)\( \rightarrow  \) 0,\( \epsilon  \)\( \rightarrow  \)
0,and k\( _{0} \) \( \rightarrow  \) \( \infty  \) limits and show which
are important to maintain. 

In this section,we wish to show several points, which we shall first enumerate
below:

(1) For \( \lambda  \)\( \neq  \)0,\( \epsilon  \)\( \neq  \)0; the propagator
has the correct high energy behavior {[} i.e. compatible with the damping behavior
in the Lorentz gauges{]} ;in other words as k\( _{0} \) \( \rightarrow  \)
\( \infty  \), the propagator behaves as k\( _{0}^{-2} \) .We have to keep
\( \lambda  \)\( \neq  \)0 for this, though of course we can take the limit
\( \lambda  \)\( \rightarrow  \) 0 in the end of the calculation. 

(2) The Coulomb gauge Feynman integrals are not well-defined except as the limit
\( \lambda  \)\( \rightarrow  \) 0. In other words, the singular gauge \( \nabla .\mathbf{A}=0 \)
is defined only as this limit.This is very similar to the light-cone gauge in
this path integral treatment {[}13,18{]}.

(3) If we do take \( \epsilon  \)\( \rightarrow  \) 0 limit first, we do recover
the naive Coulomb gauge propagator; which however does not have the desirable
high energy behavior.

(4) The discussion of high energy behavior k\( _{0} \) \( \rightarrow  \)
\( \infty  \), and the limits \( \lambda  \)\( \rightarrow  \) 0 and \( \epsilon  \)\( \rightarrow  \)
0 are all interlinked.There are terms in the propagator whose high energy behavior
crucially depends on the order of the limits: This happens because the expansion
of the logarithms \( ln \){[}\( \frac{(1-\kappa _{0})^{2}-\Delta ^{2}}{(1-a_{1})^{2}} \){]}
and \( ln \){[}1+ \( \frac{2\Delta }{\kappa _{2}(1-\kappa _{1})} \){]} (3.8)
fails for large enough k\( _{0} \), if \( \epsilon  \) and \( \lambda  \)
are fixed and positive.

(5)We shall indeed be able to retrieve the two loop anomalous Coulomb interaction
contribution obtained in the example mentioned in Ref.{[}5{]}. But in that treatment,an
extra ad hoc regularization was required. In our treatment, this term appears
naturally without an extra regularization.

Our results bring out a need for the delicate treatment of the Coulomb gauge
and it is hoped that it may be able to eliminate the troubles found in earlier
works {[}8,9{]}.

To see these results, we first make a convenient tabulation of the behavior
of various factors and terms involved in the propagator.We show only the relevant
entries in the table below.

\vspace{0.5001cm}
{\centering \begin{tabular}{|c|c|c|c|c|}
\hline 
Quantity&
value as &
\( \epsilon  \)\( \rightarrow  \)0 &
k\( _{0} \)\( \rightarrow  \)\( \infty  \) &
k\( _{0} \)\( \rightarrow  \)\( \infty  \) behavior \\
\hline 
\hline 
&
\( \epsilon  \)\( \rightarrow  \)0&
behavior&
behavior&
at \( \lambda  \) = 0\\
\hline 
\hline 
\( \alpha  \)&
&
O(1)&
k\( _{0} \)\( ^{2} \)&
O(1)\\
\hline 
\hline 
a\( _{1} \)&
k\( ^{2}/k^{2}_{0} \)&
O(1)&
O(1)&
O(1)\\
\hline 
\( \kappa  \)\( _{0} \)&
k\( ^{2}/k^{2}_{0} \)&
O(1)&
O(1)&
O(1)\\
\hline 
a\( _{1} \)-\( \kappa  \)\( _{0} \)&
0&
O(\( \epsilon  \))&
O(k\( _{0} \)\( ^{-2} \))&
O(k\( _{0} \)\( ^{-2} \))\\
\hline 
\( \Delta  \)\( ^{2} \)&
&
O(\( \epsilon  \))&
O(\( \lambda  \)k\( _{0} \)\( ^{-2} \))&
O(k\( _{0} \)\( ^{-4} \))\\
\hline 
1-a\( _{1} \)&
\textbf{|k|\( ^{2}/k^{2}_{0} \)}&
O(1)&
O(k\( _{0} \)\( ^{-2} \))&
O(k\( _{0} \)\( ^{-2} \))\\
\hline 
1-\( \kappa  \)\( _{0} \)&
\textbf{|k|\( ^{2}/k^{2}_{0} \)}&
O(1)&
O(k\( _{0} \)\( ^{-2} \))&
O(k\( _{0} \)\( ^{-2} \))\\
\hline 
(\( \kappa  \)\( _{0} \)/a\( _{1} \))\( ^{2} \)&
1&
O(1)&
O(1)&
O(1)\\
\hline 
1/\( \kappa  \)\( _{1} \)(1-\( \kappa  \)\( _{2} \))&
1/{[}\( \kappa  \)\( _{0} \)(1-\( \kappa  \)\( _{0} \))-\( \Delta  \)+\( \Delta  \)\( ^{2} \){]}&
O(1)&
O(1)&
O(1)\\
\hline 
\( ln \){[}1+ \( \frac{2\Delta }{\kappa _{2}(1-\kappa _{1})} \){]}&
\( \frac{2\Delta }{\kappa _{2}(1-\kappa _{1})} \)&
O(\( \sqrt{\varepsilon } \))&
O(1)&
O(1)\\
\hline 
\( ln \){[}\( \frac{\kappa ^{2}_{0}-\Delta ^{2}}{a^{2}_{1}} \){]}&
\( \frac{-2i\varepsilon }{k^{2}_{0}+i\varepsilon } \)-\( \frac{\Delta ^{2}}{a^{2}_{1}} \)&
O(\( \epsilon  \) )&
O(\( \epsilon  \) k\( _{0} \)\( ^{-2} \))&
O(k\( _{0} \)\( ^{-2} \))\\
\hline 
\hline 
\( ln \){[}\( \frac{(1-\kappa _{0})^{2}-\Delta ^{2}}{(1-a_{1})^{2}} \){]}&
\( \frac{(2-\kappa _{0}-a_{1})(a_{1}-\kappa _{0})-\Delta ^{2}}{(1-a_{1})^{2}} \)&
O(\( \epsilon  \) k\( _{0} \)\( ^{2} \))&
\( ln \) k\( _{0} \)\( ^{2} \)&
\( ln \) {[}k\( _{0} \)\( ^{0} \){]}\\
\hline 
\hline 
\( \frac{1}{2\Delta } \){[}(a\( _{1} \)-\( \kappa  \)\( _{0} \))(\( \alpha  \)+\( \kappa  \)\( _{0} \))
+ \( \Delta  \)\( ^{2} \){]}&
&
&
\( k_{0}^{-1} \)&
\( k_{0}^{0} \)\\
\hline 
\hline 
1/\( D \)\( \equiv  \)1/{[}(a\( _{1} \)-\( \kappa  \)\( _{0} \))\( ^{2} \)-\( \Delta  \)\( ^{2} \){]}&
-1/\( \Delta  \)\( ^{2} \)&
O(\( \epsilon  \)\( ^{-1} \))&
O(k\( _{0} \)\( ^{2} \))&
O(k\( _{0} \)\( ^{4} \))\\
\hline 
\end{tabular}\par}
\vspace{0.5001cm}

We shall now show, in sequence, how the above points arise in sequence.

(1)The above table will enable one to verify that as k\( _{0} \)\( \rightarrow  \)
\( \infty  \), every term in (3.8) except two go as k\( _{0} \)\( ^{-2} \)
{[}upto logarithms{]}. There are two exceptional terms:(a) one which goes as
O(k\( _{0} \)\( ^{0} \)\( ln \)k\( _{0} \)\( ^{2} \) ) and arises from
\( \frac{\alpha }{2} \)\( ln \){[}\( \frac{(1-\kappa _{0})^{2}-\Delta ^{2}}{(1-a_{1})^{2}} \){]}
and is proportional to -\( \frac{-\lambda (k_{0}^{2}+i\varepsilon )}{4|\mathbf{k}|^{2}} \)
\( ln \){[}\( \frac{(1-\kappa _{0})^{2}-\Delta ^{2}}{(1-a_{1})^{2}} \){]}
(b) another which goes as k\( _{0} \)\( ^{-1} \) and arises from \( \frac{\alpha }{2\Delta } \)\( ln \){[}1+
\( \frac{2\Delta }{\kappa _{2}(1-\kappa _{1})} \){]}. In most of the other
terms except one , the logarithms can be expanded and terms of O (\( \epsilon  \))
can be dropped. We deal with these terms in Appendices A,B and C.

(2) We also note that had we put \( \lambda  \) = 0 from the beginning, the
k\( _{0} \) \( \rightarrow  \) \( \infty  \) behavior of 1/\( D \)\( \equiv  \)1/{[}(a\( _{1} \)-\( \kappa  \)\( _{0} \))\( ^{2} \)-\( \Delta  \)\( ^{2} \){]}
would have been modified {[} see table{]} from O(k\( _{0} \)\( ^{2} \)) to
O(k\( _{0} \)\( ^{4} \)) . This makes the propagator behave as O(k\( _{0} \)\( ^{0} \)
). 

(3) is easily verified by direct calculation and will follow as a particular
result later.

(4) is seen, for example, by examining the behavior of 

\( \frac{1}{\varepsilon } \)ln{[}\( \frac{(1-\kappa _{0})^{2}-\Delta ^{2}}{(1-a_{1})^{2}} \){]}
=  \( \frac{1}{\varepsilon } \)ln{[}\( \frac{k_{0}^{2}}{(k_{0}^{2}+i\varepsilon )} \){]}
+ \( \frac{1}{\varepsilon } \)ln{[} A + \( i\varepsilon \lambda B \) k\( _{0} \)\( ^{2} \){]} 

with,

A\( \equiv  \) \( \frac{|\mathbf{k}|^{2}-i\varepsilon \lambda }{|\mathbf{k}|^{2}-i\varepsilon } \);B
= \( \frac{1}{(|\mathbf{k}|^{2}-i\varepsilon )^{2}} \)

We note that the expansion of \( \frac{1}{\varepsilon } \) \( ln \){[} A +
iB\( \lambda  \)\( \epsilon  \) k\( _{0} \)\( ^{2} \){]} is valid only over
the range |\( \frac{B}{A} \)\( \lambda  \)\( \epsilon  \) k\( _{0} \)\( ^{2} \)|
< 1.This range intrinsically depends on \( \epsilon  \) and \( \lambda  \).
If \( \epsilon  \) \( \rightarrow  \) 0 were to be taken first, the range
would extend to \( -\infty  \) < k\( _{0} \) < \( \infty  \) ; and the \( \frac{1}{\varepsilon } \)
\( ln \){[} A + iB\( \lambda  \)\( \epsilon  \) k\( _{0} \)\( ^{2} \){]}
would behave as k\( _{0} \)\( ^{2} \) giving rise to the badly behaved term
in the Coulomb propagator.The same is true for \( \lambda  \)\( \rightarrow  \)
0 limit. But, we always understand that \( \epsilon  \) \( \rightarrow  \)
0 is to be taken after the Feynman diagrams are calculated.\emph{If we also
define the Coulomb gauge as the} \( \lambda  \)\( \rightarrow  \) 0 \emph{limit,
then the behavior of this factor as} k\( _{0} \)\( \rightarrow  \) \( \infty  \)
is seen to be O(k\( _{0} \)\( ^{0} \) ) upto logarithms.

(5) is easily seen from the result (4.8) below if one makes use of the earlier
results of Doust {[}8{]}.This is commented upon following (4.8).

We shall now give the results for the various terms in (3.8).We shall denote
the three logarithmic terms as \{1\},\{2\},\{3\}. In the following, we assume
that |\textbf{k| \( \neq  \)}0, \( k_{0} \)\textbf{\( \neq  \)}0 and we have
chosen \( \epsilon  \) <\textcompwordmark{}< |\textbf{k|} , \( k_{0} \) .

For \{1\}, we write : 

ln{[}\( \frac{(1-\kappa _{0})^{2}-\Delta ^{2}}{(1-a_{1})^{2}} \){]} =  ln{[}\( \frac{k_{0}^{2}}{(k_{0}^{2}+i\varepsilon )} \){]}
+ ln{[} A + \( i\varepsilon \lambda B \) k\( _{0} \)\( ^{2} \){]}~~ ~~
~~~ ~~ ~~~~ ~~ ~~~ ~~(4.1)

Then \{1\} receives a net contribution from ln {[}\( \frac{k_{0}^{2}}{(k_{0}^{2}+i\varepsilon )} \){]}
factor as : 

\( \frac{|\mathbf{k}|^{2}-\lambda k_{0}^{2}}{(k^{2}+i\varepsilon )\Omega } \)~
~~ ~~~ ~~~ ~~ ~~~ ~~~~ ~~~ ~~ ~~ ~~~ ~~(4.2a), 

with 

\( \Omega  \) \( \equiv  \) \{|\textbf{k|\( ^{2} \)- \( \lambda  \)}k\( _{0} \)\( ^{2} \)-i\( \epsilon  \)\}
~~~ ~~~~~~~~ ~~~~~ ~~(4.3),

We further find additional contributions to \{1\}:

(a) \( \frac{-(1-\lambda )^{2}k_{0}^{2}}{(k^{2}+i\varepsilon )\Omega } \) ~~
~~ ~~~ ~~ ~~ ~~ ~~~ ~~ ~~ ~~~ ~~(4.2b),

(b)\( \frac{\lambda }{i\varepsilon } \)\( \frac{k_{0}^{2}}{\Omega } \) \( ln \){[}
A + \( i\varepsilon \lambda B \) k\( _{0} \)\( ^{2} \){]}~~ ~~ ~~ ~~~
~ ~~ ~~ ~ ~~ ~~~ ~~(4.2c)

= \( \frac{\lambda }{i\varepsilon } \)\( \frac{k_{0}^{2}}{|\mathbf{k}|^{2}-\lambda k_{0}^{2}-i\varepsilon } \)
\( ln \){[} A + \( i\varepsilon \lambda B \) k\( _{0} \)\( ^{2} \){]}

= -\( \frac{1}{i\varepsilon } \) \( ln \){[} A + \( i\varepsilon \lambda B \)
k\( _{0} \)\( ^{2} \){]}+\( \frac{1}{i\varepsilon } \)\( \frac{|\mathbf{k}|^{2}-i\varepsilon }{|\mathbf{k}|^{2}-\lambda k_{0}^{2}-i\varepsilon } \)
\( ln \){[} A + \( i\varepsilon \lambda B \) k\( _{0} \)\( ^{2} \){]}

= -\( \frac{1}{i\varepsilon } \) \( ln \){[} A + \( i\varepsilon \lambda B \)
k\( _{0} \)\( ^{2} \){]}+\( \frac{1}{i\varepsilon } \)\( \frac{|\mathbf{k}|^{2}-i\varepsilon }{|\mathbf{k}|^{2}-\lambda k_{0}^{2}-i\varepsilon } \)
\( ln \)A+\( \frac{1}{i\varepsilon } \)\( \frac{|\mathbf{k}|^{2}-i\varepsilon }{|\mathbf{k}|^{2}-\lambda k_{0}^{2}-i\varepsilon } \)
\( ln \){[} 1 + \( i\varepsilon \lambda \frac{B}{A} \) k\( _{0} \)\( ^{2} \){]}

\( \equiv  \) -(\( \Re _{1} \) +\( \Re _{2} \)+ \( \Re _{3} \))~ ~~ ~~~
~~ ~~~ ~~~ ~~~ ~~~~ ~~~ ~~(4.3)

and 

(c) \( \frac{\lambda -1}{i\varepsilon } \)\( \frac{|\mathbf{k}|^{2}k_{0}^{2}}{(k^{2}+i\varepsilon )(|\mathbf{k}|^{2}-\lambda k_{0}^{2}-i\varepsilon )} \)
\( ln \){[} 1 + \( i\varepsilon \lambda \frac{B}{A} \) k\( _{0} \)\( ^{2} \){]}\( \equiv  \)-
\( \Re _{4} \)~ ~~~ ~~~~ ~~~ ~~(4.2d)

\( \Re _{1} \) ,\( \Re _{2} \), \( \Re _{3} \) and \( \Re _{4} \) are dealt
with in appendix A. We find the contributions \{2\} and \{3\} as

\{2\} =  \( \frac{2(\lambda k_{0}^{2}-|\mathbf{k}|^{2})}{(k^{2}+i\varepsilon )\Omega } \)
+ \( \frac{(\lambda k_{0}^{2}-|\mathbf{k}|^{2})}{(k^{2}+i\varepsilon )^{2}} \)
~~~~ ~~~ ~~~~~ ~~~ ~~(4.4).

\{3\} =  \( \frac{2k_{0}^{2}}{(k^{2}+i\varepsilon )\Omega } \) - \( \Re _{5} \)
~~ ~~~ ~~~~ ~~~ ~~~~ ~~~ ~~~~ ~~~ ~~(4.5)

\( \Re _{5} \) is dealt with in appendix B.

We add the result to find:

G\( ^{0C} \)\( _{00} \)(\( k \))-G\( ^{0L} \)\( _{00} \)(\( k \)) = \( \frac{(\lambda k_{0}^{2}-|\mathbf{k}|^{2})}{(k^{2}+i\varepsilon )^{2}} \)-\( \frac{(-1-3\lambda +\lambda ^{2})k_{0}^{2}+|\mathbf{k}|^{2}}{(k^{2}+i\varepsilon )(|\mathbf{k}|^{2}-\lambda k_{0}^{2}-i\varepsilon )} \)
- \( \sum  \) \( ^{5} \)\( _{_{i=1}} \)\( \Re _{i} \)~ ~~~~ ~ ~~(4.6)

We shall show, in appendices A and B, that the net effect of the residual terms
\( \Re _{i} \) ; \( i \) =  1,...,5 to the coordinate space propagator is
given by the first term in (A.13) . Incorporating this effect, we find\footnote{%
We emphasize that this propagator is \emph{not} the naive propagator \emph{in}
the Coulomb gauge with gauge parameter \( \lambda  \) : the latter would equal
to G\( _{\mu \nu } \)\( ^{0M} \) of (3.3) at \( \kappa  \)=1 .
}

G\( ^{0C} \)\( _{00} \)(\( k \))-G\( ^{0L} \)\( _{00} \)(\( k \)) = \( \frac{(\lambda k_{0}^{2}-|\mathbf{k}|^{2})}{(k^{2}+i\varepsilon )^{2}} \)+\( \frac{(1+4\lambda -\lambda ^{2})k_{0}^{2}-|\mathbf{k}|^{2}}{(k^{2}+i\varepsilon )(|\mathbf{k}|^{2}-\lambda k_{0}^{2}-i\varepsilon )} \)~~~
~~~ ~~(4.7)

We note, as a check, that at \( \lambda  \) = \( \epsilon  \) = 0, the right
hand side reduces to -\( \frac{k_{0}^{2}(k_{0}^{2}-2|\mathbf{k}|^{2})}{k^{4}|\mathbf{k}|^{2}} \)
as is expected for this (naive) propagator difference. We further note that
G\( ^{0C} \)\( _{00} \)(\( k \)) above behaves as \( k_{0}^{-2} \) \textbf{\emph{if
we keep \( \lambda  \)\( \neq  \)0.}} If we were to first put \textbf{\emph{\( \lambda  \)\( = \)0;}}
G\( ^{0C} \)\( _{00} \)(\( k \)) would behave as \( k_{0}^{0} \) spoiling
the ultraviolet behavior of some Feynman integrals as compared to the Lorentz
gauges.Therefore, we believe that one must do integrals with the propagator
given by (4.7) \emph{with \( \lambda  \)\( \neq  \)0 and take the limit \( \lambda  \)\( \rightarrow  \)0
only at the end}\textbf{\emph{.}} We further note that there is no need to invoke
any \emph{additional regularization} in this formalism which is a path-integral
formalism. This is seen as the root of any non-trivial effects arising in the
Coulomb gauge in this formalism.

Substituting for G\( ^{0L} \)\( _{00} \)(\( k \)):

G\( ^{0L} \)\( _{00} \)(\( k \)) =  \( \frac{-1}{k^{2}+i\varepsilon } \)\{
g\( _{00} \) -\( \frac{k_{0}^{2}(1-\lambda )}{k^{2}+i\varepsilon \lambda } \)\},

we find 

G\( ^{0C} \)\( _{00} \)(\( k \)) = \( \frac{1}{(|\mathbf{k}|^{2}-\lambda k_{0}^{2}-i\varepsilon )} \)+\( \frac{(4\lambda -\lambda ^{2})k_{0}^{2}}{(k^{2}+i\varepsilon )(|\mathbf{k}|^{2}-\lambda k_{0}^{2}-i\varepsilon )} \)+
O(\( \epsilon  \))~~~ ~~~ ~~(4.8)

We shall provide a comparison with the procedure of Doust {[}8{]} in the next
section. But we note that the first term in (4.8) has identical appearance as
the propagator of Doust with the replacement \( \theta  \)\( ^{2} \)\( \rightarrow  \)\( \lambda  \).Thus,
as far as the two loop integral in ref. {[}5{]} is concerned,the results of
Doust are sufficient to imply that the results of {[}5{]} are reproduced by
this term.The second term in (4.8) does not contribute to this integral as \( \lambda  \)
\( \rightarrow  \) 0.

\section{\noun{Conclusions, Comparison with Other Works and Additional Comments}}

We summarize the results we have obtained. We started out with the path-integral
construction for an arbitrary gauge that is compatible with the Lorentz gauges
and which has been applied to the problems of the axial and planar gauges. We
followed a procedure similar to the one used for the axial gauges to obtain
the Coulomb propagator.We found that unlike the naive Coulomb propagator, the
propagator so obtained has a good high energy behavior at high energies except
for a set of terms \( \Re  \)\( _{i} \) 's. We then obtained an effective
treatment for these terms following the earlier work for the axial gauges by
requiring that effective terms contribute to the co-ordinate space propagator
the same way as these terms in the limit \( \lambda  \),\( \epsilon  \) \( \rightarrow  \)
0.We found that the resultant propagator is well-behaved as a function of k\( _{0} \).
It reproduces a term of the form used by Doust {[}8{]} in his interpolating
gauge treatment , but with a modified interpretation.We also find an extra term
which formally vanishes at \( \lambda  \) = 0 but which is \( \lambda  \)
multiplied by a term which is O(k\( _{0} \)\( ^{0} \)) at \( \lambda  \)
=  0 and it can possibly contribute to sufficiently higher order diagrams.

We shall provide a comparison with the procedure of Doust {[}8{]} as it has
resemblance to ours in appearance. Doust has written down the Coulomb propagator
that has resemblance to the first term in (4.8) with his \( \theta  \)\( ^{2} \)\( \rightarrow  \)
\( \lambda  \) . We first point out the differences :

(a) The propagator we have obtained is in the {}``family{}'' of the Coulomb
gauges with \( \lambda  \) the gauge parameter.The singular Coulomb gauge is
understood as the \( \lambda  \) \( \rightarrow  \) 0. It is this gauge parameter
that appears to regulate the behavior of the time-like propagator as opposed
to the parameter \( \theta  \) of {[}8{]} that takes one away from the {[}{}``family{}''
of{]} Coulomb gauge .

(b)We find additional terms in the Coulomb propagator as shown in (4.8). Whether
these terms can in fact contribute in sufficiently higher orders even as \( \lambda  \)
\( \rightarrow  \) 0 , and in particular, they remove the problems mentioned
by Doust and Taylor {[}8{]} requires further study . 

(c) As emphasized earlier in Sec. 2A , we have to treat the \( \epsilon  \)-term
in the path-integral carefully if we are to preserve the gauge-independence
of the vacuum expectation value of a gauge-invariant observable O{[}A{]}. The
present treatment indeed is built to take care of this; we have reservations
if the treatment of Doust in fact preserves the gauge-independence as \( \theta  \)
is varied {[}22{]} .

We make further comparisons with recent works{[}11,12{]}:

(a)Our formulation of the Coulomb gauge proceeds from a path-integral formalism
that in effect includes the procedure for making the energy integrals regularized.
No further regularization is necessary.

(b) WT identities can be written directly in this formulation which is exact
in the sense that \emph{no additional interpretation} is imposed on the Green's
functions in it from outside.

(c) Equivalence of this path-integral for the Coulomb gauges with that for the
Lorentz gauges \{at least for the vacuum expectation values of gauge-invariant
operators\} is implicit in its construction. This does not seem obvious in other
attempts.

\textbf{\emph{\noun{Acknowledgements}}}

One of us {[}SDJ{]} would like to acknowledge the support of Department of Science
and Technology, India {[}grant no. DST/PHY/-19990170{]}. BPM would like to acknowledge
hospitality of Dept. of Physics, I.I.T.Kanpur during a visit.

\section{Appendix A}

In this appendix, we shall develop,along the lines of {[}13{]}, an effective
treatment for the residual terms \( \Re _{i} \) (\( i=1,....4) \) mentioned
in Sec. 4. The basic idea is to replace it by a term that gives the same contribution
to the co-ordinate space propagator as \( \lambda  \) and/or \( \epsilon  \)\( \rightarrow  \)
0 . We begin with

\( \Re _{1} \) = \( \frac{1}{i\varepsilon } \) \( ln \){[} A + \( i\varepsilon \lambda B \)
k\( _{0} \)\( ^{2} \){]}\( \qquad \qquad \qquad \qquad \qquad \qquad \qquad \qquad \qquad  \)(A.1)

and consider the Fourier transform  

\( \frac{1}{i\varepsilon } \)\( \int  \)d\( k_{0} \)exp(\( -ik_{0} \)t)\( ln \){[}
A + \( i\varepsilon \lambda B \) k\( _{0} \)\( ^{2} \){]}\( \qquad \qquad \qquad \qquad \qquad \qquad \qquad \qquad \qquad  \)(A.2)

In defining the Fourier transform, it is always understood that a convergence
factor such as {\small \( \begin{array}{c}
lim\\
\delta \rightarrow 0
\end{array} \)} exp(\( -|k_{0}|\delta  \)) is used. We integrate (A.2) by parts to find,

\( \Re _{1} \)(t) =  -\( \frac{i}{\varepsilon } \)exp(\( -ik_{0} \)t\( -|k_{0}|\delta  \))\( ln \){[}
A + \( i\varepsilon \lambda B \) k\( _{0} \)\( ^{2} \){]}\( \frac{1}{-it-\delta } \)\( \mid ^{^{\infty }}_{_{0}} \)-\( \frac{i}{\varepsilon } \)exp(\( -ik_{0} \)t\( -|k_{0}|\delta  \))\( ln \){[}
A + \( i\varepsilon \lambda B \) k\( _{0} \)\( ^{2} \){]}\( \frac{1}{-it+\delta } \)\( \mid ^{^{0}}_{_{-\infty }} \)-\( \frac{i}{\varepsilon t} \)\( \int  \)d\( k_{0} \)exp(\( -ik_{0} \)t)\( \frac{2B\lambda \varepsilon k_{0}}{A+iB\lambda \varepsilon k^{2}_{0}} \)\( \qquad \qquad \qquad  \)\( \qquad \qquad  \)(A.3)

The first two terms lead to a contribution -\( \frac{i}{\varepsilon } \)\( ln \)
A {\small \( \begin{array}{c}
lim\\
\delta \rightarrow 0
\end{array} \)} \{\( \frac{1}{it+\delta } \)+\( \frac{1}{-it+\delta } \)\} =  -\( \frac{i\pi }{\varepsilon } \)\( ln \)
A \( \delta  \)(t)

The last integral is, for t\( \neq  \)0:

\( \frac{2\pi i}{\varepsilon t} \)exp ( - \( \sqrt{\frac{A}{2B\lambda \varepsilon }} \)
|t|) {[} \( \theta  \)(t)exp ( i\( \sqrt{\frac{A}{2B\lambda \varepsilon }} \)
t)-\( \theta  \)(-t)exp (- i\( \sqrt{\frac{A}{2B\lambda \varepsilon }} \)
t){]} \( \qquad \qquad \qquad  \)(A.4)

We note that for t\( \neq  \)0, the above vanishes as \( \lambda  \) \( \rightarrow  \)
0\( ^{+} \). Thus both the terms have a support only at t = 0 as \( \lambda  \)
\( \rightarrow  \) 0\( ^{+} \).

We shall now consider \( \Re _{2} \).

\( \Re _{2} \) =  -\( \frac{1}{i\varepsilon } \)\( \frac{|\mathbf{k}|^{2}-i\varepsilon }{|\mathbf{k}|^{2}-\lambda k_{0}^{2}-i\varepsilon } \)
\( ln \)A =  \( \frac{1-\lambda }{|\mathbf{k}|^{2}-\lambda k_{0}^{2}-i\varepsilon } \)
\( \qquad \qquad \qquad  \)(A.5)

The Fourier transform reads:

\( \Re _{2} \)(t) =  \( \int  \) \( \frac{(1-\lambda )exp(-ik_{0}t)}{|\mathbf{k}|^{2}-\lambda k_{0}^{2}-i\varepsilon } \)dk\( _{0} \)\( \qquad \qquad \qquad  \)\( \qquad \qquad \qquad  \)(A.6)

For t > 0, we can close the contour below to find:

\( \Re _{2} \)(t) =  2\( \pi  \)i\( \frac{1-\lambda }{\lambda } \)\( \frac{exp(-iKt)}{K} \)
= 2\( \pi  \)i\( \frac{1-\lambda }{\sqrt{\lambda }} \)\( \frac{exp(-iKt)}{|\mathbf{k}|} \);
K =  \( \frac{|\mathbf{k}|}{\sqrt{\lambda }} \)\( \qquad \qquad \qquad  \)(A.7)

As \( \lambda  \) \( \rightarrow  \) 0, the exponential oscillates rapidly
for t \( \neq  \)0. Hence, we can replace the expression by its average over
a cycle \( \Delta  \)t = \( \frac{2\pi \sqrt{\lambda }}{|\mathbf{k}|} \) <
< t for sufficiently small \( \lambda  \). This average is zero. A similar
argument can be given for t < 0.Thus \( \Re _{2} \)(t) has support only at
t = 0.

{[} Alternately, we may include a convergence factor {\small \( \begin{array}{c}
lim\\
\delta \rightarrow 0
\end{array} \)} exp(\( -k_{0}^{2}\delta  \)) in the definition of the Fourier transform to
deal with the term.{]}

To find the net contribution from \( \Re _{1} \)(t)+\( \Re _{2} \)(t), we
consider 

\( \int ^{\infty }_{_{-\infty }} \) \{\( \Re _{1} \)(t)+\( \Re _{2} \)(t)\}
dt =  2\( \pi  \) \{\( \Re _{1} \)(k\( _{0} \))+\( \Re _{2} \)(k\( _{0} \))\}\( \mid _{_{k_{0}=0}} \)
\( \qquad \qquad \qquad  \)(A.8)

= \( \frac{1}{i\varepsilon } \) \( ln \)A-\( \frac{1}{i\varepsilon } \) \( ln \)A
=  0.\( \qquad \qquad \qquad  \)\( \qquad \qquad \qquad  \)(A.9)

Thus, the two terms do not contribute to the propagator in the sense described.

Next we shall deal with the effective treatment of \( \Re _{3} \)(k\( _{0} \))+\( \Re _{4} \)(k\( _{0} \)).

\( \Re _{3} \)(k\( _{0} \))+\( \Re _{4} \)(k\( _{0} \)) =  -\( \frac{1}{i\varepsilon } \)\( \frac{|\mathbf{k}|^{2}-i\varepsilon }{|\mathbf{k}|^{2}-\lambda k_{0}^{2}-i\varepsilon } \)
\( ln \){[} 1 + \( i\varepsilon \lambda \frac{B}{A} \) k\( _{0} \)\( ^{2} \){]} 

\( \qquad \qquad \qquad  \)+ \( \frac{1-\lambda }{i\varepsilon } \)\( \frac{|\mathbf{k}|^{2}k_{0}^{2}}{(k^{2}+i\varepsilon )(|\mathbf{k}|^{2}-\lambda k_{0}^{2}-i\varepsilon )} \)
\( ln \){[} 1 + \( i\varepsilon \lambda \frac{B}{A} \) k\( _{0} \)\( ^{2} \){]}.\( \qquad \qquad \qquad  \)(A.10)

{[} We note that we cannot carry out an expansion of the logarithm in these
terms as the condition for the expansion of their validity |\( i\varepsilon \lambda \frac{B}{A} \)
k\( _{0} \)\( ^{2} \)| <1 is not fulfilled nor are these terms compatible
with the weaker conditions in the appendix C. Were an expansion to be carried
out, this will spoil the large k\( _{0} \)-behavior. The actual large k\( _{0} \)-behavior
is as k\( _{0} \)\( ^{-2} \)ln k\( _{0} \), whereas on expansion, it will
yield O{[}1{]} terms and this is because of an invalid expansion {]}. We however
find that when we combine such terms, we find

\( \Re _{3} \)(k\( _{0} \))+\( \Re _{4} \)(k\( _{0} \)) = \( \frac{1}{i\varepsilon } \)\( \frac{|\mathbf{k}|^{2}}{\{|\mathbf{k}|^{2}-\lambda k_{0}^{2}-i\varepsilon \}(k^{2}+i\varepsilon )} \)
\( ln \){[} 1 + \( i\varepsilon \lambda \frac{B}{A} \) k\( _{0} \)\( ^{2} \){]}

\( \qquad \qquad \qquad  \)\( \bullet  \)\{ (1-\( \lambda  \)) k\( _{0}^{2} \)
- (k\( ^{2}+i\varepsilon ) \)\( \left[ 1-\frac{i\varepsilon }{|\mathbf{k}|^{2}}\right]  \)\}\( \qquad \qquad \qquad  \)(A.11)

In the above combination, the leading term at \( \lambda  \) =  0 cancels.
We find,

= \( \frac{1}{i\varepsilon } \)\( \frac{-|\mathbf{k}|^{2}\lambda k_{0}^{2}}{\{|\mathbf{k}|^{2}-\lambda k_{0}^{2}-i\varepsilon \}(k^{2}+i\varepsilon )} \)
\( ln \){[} 1 + \( i\varepsilon \lambda \frac{B}{A} \) k\( _{0} \)\( ^{2} \){]}

+\( \qquad \qquad  \)\( \frac{1}{i\varepsilon } \) \( \frac{|\mathbf{k}|^{4}}{\{|\mathbf{k}|^{2}-\lambda k_{0}^{2}-i\varepsilon \}(k^{2}+i\varepsilon )} \)
\( ln \){[} 1 + \( i\varepsilon \lambda \frac{B}{A} \) k\( _{0} \)\( ^{2} \){]}
+ O (\( \epsilon  \))\( \qquad \qquad \qquad  \)(A.12)

In the second term, we can expand the \( ln \){[} 1 + \( i\varepsilon \lambda \frac{B}{A} \)
k\( _{0} \)\( ^{2} \){]} factor {[} See (C.4){]}.This yields,

-{[}\( \Re _{3} \)(k\( _{0} \))+\( \Re _{4} \)(k\( _{0} \)){]} =  \( \frac{\lambda k_{0}^{2}}{\{|\mathbf{k}|^{2}-\lambda k_{0}^{2}-i\varepsilon \}(k^{2}+i\varepsilon )} \) 

\( \qquad \qquad  \)\( \qquad \qquad  \)-\( \frac{1}{i\varepsilon } \) \( \frac{|\mathbf{k}|^{2}\lambda k_{0}^{2}}{\{|\mathbf{k}|^{2}-\lambda k_{0}^{2}-i\varepsilon \}(k^{2}+i\varepsilon )} \)
\( ln \){[} 1 + \( i\varepsilon \lambda \frac{B}{A} \) k\( _{0} \)\( ^{2} \){]}+
O (\( \epsilon  \)).\( \qquad \qquad  \)(A.13)

We shall show in Appendix C that the second term above does not contribute {[}See
(C.2){]} to the coordinate propagator as \( \lambda  \) \( \rightarrow  \)
0 .The first term has already been incorporated in (4 .7).

\section{Appendix \textmd{B}}

In this appendix, we shall show that the term \( \Re _{5} \) (k\( _{0}) \)
of eq.(4.5) does not contribute to the coordinate space gauge boson propagator
as \( \lambda  \)\( \rightarrow  \)0. 

\( \Re _{5} \) (k\( _{0}) \) \( \equiv  \)\( \lambda  \) \( \frac{1}{\Omega } \)\( \frac{1}{\Delta } \)ln
\( \left[ 1+\frac{2\Delta }{\kappa _{2}(1-\kappa _{1})}\right]  \)\( \qquad \qquad \qquad  \)\( \qquad \qquad \qquad  \)(B.1)

= \( \lambda  \) \( \frac{k_{0}^{3}}{\sqrt{i\varepsilon (k^{2}+\iota \varepsilon )\{|\mathbf{k}|^{2}-\lambda (k_{0}^{2}+i\varepsilon )\}}} \)\( \frac{1}{(|\mathbf{k}|^{2}-\lambda k_{0}^{2}-i\varepsilon )} \)ln
\( \left[ 1+\frac{2\Delta }{\kappa _{2}(1-\kappa _{1})}\right]  \)\( \qquad \qquad \qquad  \)(B.2)

We have already defined the branch-cuts of the two square-roots.An analysis
of the singularities shows that there are branch-points for the logarithms at
\( \pm  \)\( \sqrt{|\mathbf{k}|^{2}-i\varepsilon \lambda } \) and at \( \pm  \)\( \frac{|\mathbf{k}|^{2}}{\sqrt{i\varepsilon \lambda }} \).
We can choose the four branch-cuts so that they do not cross the real axis.The
contribution of to the coordinate propagator is 

\( \Re _{5} \)(t) =  \( \int  \)dk\( _{0} \) exp(\( -ik_{0} \)t)\( \Re _{5} \)
(k\( _{0}) \) 

= \( \int  \)dk\( _{0} \) exp(\( -ik_{0} \)t)\( \lambda  \) \( \frac{k_{0}^{3}}{\sqrt{i\varepsilon (k^{2}+\iota \varepsilon )\{|\mathbf{k}|^{2}-\lambda (k_{0}^{2}+i\varepsilon )\}}} \)\( \frac{1}{(|\mathbf{k}|^{2}-\lambda k_{0}^{2}-i\varepsilon )} \)ln
\( \left[ 1+\frac{2\Delta }{\kappa _{2}(1-\kappa _{1})}\right]  \)\( \qquad \qquad \qquad  \)(B.3)

At this stage, we may be tempted to take the limit \( \lambda  \) \( \rightarrow  \)
0 directly and set \( \Re _{5} \) (k\( _{0}) \) to zero. We should however
note that putting \( \lambda  \) \( = \) 0 inside the integral alters the
convergence properties of the integral {[} it becomes naively cubically divergent:
ln \( \left[ 1+\frac{2\Delta }{\kappa _{2}(1-\kappa _{1})}\right]  \) \( \rightarrow  \)
constant for large k\( _{0} \){]}. Hence, this procedure is ill-defined; in
fact the rigorous results in Appendix C do not support this. {[}Moreover, if
this term is found inside a divergent Feynman integral, there is no exp(\( -ik_{0} \)t)
to improve convergence{]}.

We shall interpret \( \Re _{5} \) (k\( _{0}) \) in the sense of generalized
functions or distributions {[}20 {]}. Thus the propagator in the momentum space
{[}or a term contributing to it{]} is a distribution with \emph{local value}
{[}20{]} given by the number \( \Re _{5} \) (k\( _{0}) \). In order that this
interpretation is possible, we have to be able to define at least one sequence
of {}`` good{}'' functions that in tends to \( \Re _{5} \) (k\( _{0}) \).
We note that {[}despite the fact that there are branch-cuts for \( \Re _{5} \)
(k\( _{0}) \) in the \emph{complex k\( _{0} \)}-plane{]} the function is infinitely
differentiable on the \emph{real axis} {[}for -\( \infty  \) < k\( _{0}< \)\( \infty  \)
; \emph{including k\( _{0}=0] \)} so that \{\( \Re _{5} \) (k\( _{0}) \)
exp ( - k\( _{0} \)\( ^{2} \)/n\( ^{2}) \) ; n = 1,2,.....\} constitutes
such a sequence. We note then that the Fourier transform \( \Re _{5} \) (t;
|\textbf{k|)} a distribution itself {[}20{]}.We shall find it useful to deal
with \( \Re _{5} \) (t; |\textbf{k|)} in terms of its moments:

M\( _{n} \) =  \( \int  \) t\( ^{n} \)\( \Re _{5} \) (t; |\textbf{k|)} dt.\( \qquad \qquad \qquad  \)\( \qquad \qquad \qquad  \)(B.4)

We note that as \( \Re _{5} \) (t; |\textbf{k|)} is a distribution, we have
to interpret the above as an integral of a distribution multiplied by a {}``good{}''
function {[}20{]}. We consider the good function \textbf{}t\( ^{n} \) \( \exp (-t^{2}/N) \)
and define;

M\( _{n} \) \( \equiv  \) {\small \( \begin{array}{c}
lim\\
N\rightarrow \infty 
\end{array} \)}\( \int  \) t\( ^{n} \)\( \exp (-t^{2}/N) \) \( \Re _{5} \) (t; |\textbf{k|)}
dt\( \qquad \qquad \qquad  \)(B.5)

It is easy to show that this equals,

M\( _{n} \) \( = \) {\small \( \begin{array}{c}
lim\\
N\rightarrow \infty 
\end{array} \)}\( \int  \)dk\( _{0} \)\( \Re _{5} \) (k\( _{0}) \) \( \int  \)t\( ^{n} \)\( \exp (-t^{2}/N) \)
\textbf{}exp(\( -ik_{0} \)t) dt\( \qquad \qquad \qquad  \)(B.6)

= {\small \( \begin{array}{c}
lim\\
N\rightarrow \infty 
\end{array} \)}\( \int  \)dk\( _{0} \)\( \Re _{5} \) (k\( _{0}) \)\( \left( \frac{id}{dk_{0}}\right)  \)\( ^{n} \)
\( \int  \)\( \exp (-t^{2}/N) \) \textbf{}exp(\( -ik_{0} \)t) dt\( \qquad \qquad \qquad  \)(B.7)

This is easily shown to lead to 

M\( _{n} \) =  \( \left( -i\frac{d}{dk_{0}}\right)  \)\( ^{n} \)\( \Re _{5} \)
(k\( _{0}) \)\( \Vert  \)\( _{_{_{k_{0}=0}}} \)\( \qquad \qquad \qquad  \)\( \qquad \qquad \qquad  \)(B.8)

M\( _{n} \) 's exist and vanish as \( \lambda  \) \( \rightarrow  \) 0. Thus
\( \Re _{5} \) does not contribute to the Coulomb propagator in the coordinate
space {[}defined as the Fourier transform of the momentum space propagator looked
as a distribution{]} in this sense.

\section{Appendix C}

In this appendix we shall summarize the rigorous mathematical results {[}21{]}
we need and their application to the integrals involved in \( \Re _{i} \) 's.These
results refer to the interchange of the order of a limit and an integration.These
results will give the justification for several steps used earlier. {[} In the
following, b is a constant independent of parameter \( \alpha  \){]}

\textbf{\emph{Result 1}}:Let \( \int  \)\( ^{\infty }_{_{b}} \) f (x,\( \alpha  \)
)dx be \emph{uniformly convergent} when the parameter \( \alpha  \) lies in
a domain S. Then if f (x,\( \alpha  \) ) is a continuous function of both x
and \( \alpha  \) for x \( \geq  \) b and \( \alpha  \) in S, then \( \int  \)\( ^{\infty }_{_{b}} \)
f (x,\( \alpha  \) )dx is a continuous function of \( \alpha  \).

In this case we can write,

{\small \( \begin{array}{c}
lim\\
\alpha \rightarrow 0
\end{array} \)} \( \int  \)\( ^{\infty }_{_{b}} \) f (x,\( \alpha  \) )dx =   \( \int  \)\( ^{\infty }_{_{b}} \)
f (x,0 )dx\( \qquad \qquad \qquad  \)\( \qquad \qquad \qquad  \)(C.1)

provided \( \alpha  \) =  0 \( \in  \) S.

Using this result, we can easily establish that 

{\small \( \begin{array}{c}
lim\\
\lambda \rightarrow 0
\end{array} \)} \( \int  \)dk\( _{0} \)\( \frac{1}{i\varepsilon } \) \( \frac{|\mathbf{k}|^{2}\lambda k_{0}^{2}}{\{|\mathbf{k}|^{2}-\lambda k_{0}^{2}-i\varepsilon \}(k^{2}+i\varepsilon )} \)
\( ln \){[} 1 + \( i\varepsilon \lambda \frac{B}{A} \) k\( _{0} \)\( ^{2} \){]}
=  0\( \qquad \qquad \qquad  \)(C.2)

\textbf{\emph{Result 2}}: The equation \( \frac{d}{d\alpha } \)\( \int  \)\( ^{\infty }_{_{b}} \)
f (x,\( \alpha  \) )dx =  \( \int  \)\( ^{\infty }_{_{b}} \)\( \frac{d}{d\alpha } \)
f (x,\( \alpha  \) )dx holds if the integral on the right converges uniformly
and the integrand \( \frac{d}{d\alpha } \) f (x,\( \alpha  \) ) is a continuous
function of both variables when x\( \geq  \) b and \( \alpha  \) \( \in  \)S
and the integral \( \int  \) \( ^{\infty }_{_{b}} \) f (x,\( \alpha  \))
dx is convergent.

As a consequence,we have 

{\small \( \begin{array}{c}
lim\\
\varepsilon \rightarrow 0
\end{array} \)}\( \frac{1}{\varepsilon } \)\{\( \int  \)\( ^{\infty }_{_{b}} \) f (x,\( \varepsilon  \)
)dx - \( \int  \)\( ^{\infty }_{_{b}} \) f (x,0 )dx\} =  \( \int  \)\( ^{\infty }_{_{b}} \)\( \frac{d}{d\alpha } \)
f (x,\( \alpha  \) )\( \mid _{_{\alpha =0}} \) dx\( \qquad \qquad \qquad  \)(C.3)

We apply this result to justify the expansions of the logarithmic factors where
possible.Thus, for example, let F{[}k\( _{0} \),\( \epsilon  \){]} be a function
with the properties F{[}k\( _{0} \),\( \epsilon  \){]} \( \sim  \) k\( _{0} \)\( ^{-4} \)
(upto logarithms) as |k\( _{0} \)| \( \rightarrow  \)\( \infty  \) and such
that F{[}k\( _{0} \),\( \epsilon  \) = 0{]} = 0 . Then we can easily show
that

{\small \( \begin{array}{c}
lim\\
\varepsilon \rightarrow 0
\end{array} \)}\( \frac{1}{\varepsilon } \)\( \int  \)F{[}k\( _{0} \),\( \epsilon  \){]}dk\( _{0} \)
= \( \int  \)\( \left. \frac{dF[k_{0},\varepsilon ]}{d\varepsilon }\right|  \)\( _{_{_{\varepsilon =0}}} \)dk\( _{0} \)\( \qquad \qquad \qquad  \)(C.4)

This relation has been used in justifying the expansion of the logarithms where
it has been applied. (4.2c) is an example of a term where such an expansion
has been used; (4.2d) is, on the other hand, an example of a term where such
an expansion is not allowed.

\end{document}